\documentclass[journal]{IEEEtran}

\usepackage{amsmath,mathtools}
\usepackage{graphicx}
\graphicspath{{pictures/}{./}}
\usepackage{amsthm}
\usepackage{amsfonts}
\usepackage{cleveref}
\usepackage{array}
\usepackage{xcolor}
\usepackage{epstopdf}
\usepackage{float}

\begin{document}

\title{3D User Localization for Planar Arrays in LoS Near- and Far-Fields via Summed Phase Differences}

\author{Sergey Isaev and Nikola~Zlatanov%
\thanks{S.~Isaev and N.~Zlatanov are with Innopolis University, Innopolis, 420500, Russia (e-mails: s.isaev@innopolis.ru and n.zlatanov@innopolis.ru).}
}

\maketitle

\begin{abstract}
This letter presents a phase-difference-based scheme for three-dimensional (3D) line-of-sight (LoS) user localization using a uniform planar array (UPA), applicable to both near-field and far-field regimes under the exact spherical-wave model. Unlike the previously studied two-dimensional (2D) uniform linear array (ULA) case, the 3D UPA case requires jointly exploiting the two array axes in order to recover the user's range, azimuth, and zenith angle. Adjacent-antenna phase-differences are first estimated from uplink pilots and then summed along the array axes to obtain unwrapped phase-differences between widely separated antenna elements. These summed phase-differences enable the construction of multiple three-equation systems whose solutions yield the user's range, azimuth, and zenith angle. We quantify the number of such equation systems, provide a representative closed-form estimator that uses only three phase-difference sums, and propose an all-data nonlinear least-squares estimator that exploits all available sums. Numerical results include a classical Fraunhofer near-field case and absolute 3D position RMSE in meters, and show near-CRB performance when the closed-form initialization is reliable. Moreover, unlike state-of-the-art baseline schemes, whose performance depends on well-tuned hyperparameters, the proposed estimators are hyperparameter-free.
\end{abstract}


\IEEEpeerreviewmaketitle

\section{Introduction}
The deployment of transceivers equipped with extra-large antenna arrays at higher frequency bands, such as millimeter-wave (mmWave) and terahertz (THz), significantly expands the near-field region \cite{liu2023}. As a result, many users served by such arrays may operate in a regime where the spherical nature of the wavefront cannot be neglected. Whether the goal is accurate beamforming/beamfocusing or direct user positioning, the demand for high-precision localization has therefore grown substantially.

Traditional localization techniques are predominantly based on far-field assumptions, where the received wavefront is approximated as planar. This simplification breaks down with electrically large arrays. In the near-field, the exact distance from the user to each antenna element must be accounted for, which leads to distance-dependent phase and amplitude variations across the array. This makes localization more challenging, but also creates additional geometric information that can be exploited if the exact spherical-wave model is used.

The existing body of work on 3D near-field localization has mainly followed two paths. The first path is based on exhaustive search over a discretized parameter grid \cite{chen2022, pan2020}. While potentially accurate, such methods can be computationally demanding for large arrays and fine grids. The second path is based on simplified near-field models, such as Fresnel approximations \cite{tao2011_fresnel}, which reduce the mathematical burden at the price of model mismatch.

Among the methods developed for the same exact spherical-wave system model considered here, the work closest to ours is \cite{baseline}. In \cite{baseline}, the uniform planar array (UPA) is partitioned into multiple subarrays such that, although the user may lie in the near-field of the full array, it can be approximated as being in the far-field of each subarray. The resulting angle-of-arrival drift across subarrays is then fused within a Bayesian message-passing framework (APLE), and an enhanced variant (E-APLE) further refines the APLE estimate by solving a near-field maximum-likelihood problem for the full array. We therefore use \cite{baseline} as the main benchmark in our numerical results.

This paper tackles the 3D localization problem directly under the exact spherical-wave model and without subarray partitioning. Our method relies on phase-differences, which eliminate the need to know the gain functions and make the estimator insensitive to unknown amplitude scaling. Most importantly, the extension from the known 2D ULA result in \cite{n.zlat} to the present 3D UPA setting is not a straightforward dimensional lift. In the 2D ULA case there is only one array axis and two geometric unknowns. In the 3D UPA case the two array axes are coupled through the same three unknown coordinates, and one has to combine horizontal and vertical phase-sum relations in a way that makes range, azimuth, and zenith angle jointly identifiable. This is the main new challenge addressed in this paper. Moreover, unlike the estimator in \cite{n.zlat}, the all-sum least-squares estimator proposed here approaches the 3D position CRB when the closed-form initialization is sufficiently reliable.

The main contributions of this paper are fourfold:
\begin{enumerate}
    \item We propose a 3D localization scheme based on phase-differences of signals received at adjacent UPA antenna elements. The scheme is derived for the exact spherical-wave model and is therefore applicable in both near-field and far-field regimes.
    \item We show how to construct multiple systems of three equations from summed phase-differences along the two array axes, each system enabling a direct estimate of the user's range, azimuth, and zenith angle. We also quantify the number of such systems as a function of the UPA size.
    \item We provide a representative closed-form estimator with a particularly simple expression, which requires only three phase-difference sums and yields a deterministic single-shot estimate with no algorithmic hyperparameter tuning.
\item We propose an all-data nonlinear least-squares estimator that exploits all available phase-difference sums at the UPA. This estimator can be solved numerically, e.g., via the Levenberg--Marquardt algorithm, and is evaluated through absolute 3D position RMSE in meters, including a scenario satisfying the classical Fraunhofer near-field condition.
\end{enumerate}

In terms of tradeoffs, the proposed closed-form estimator has very low computational load after phase-sum formation and uses only the standard single-user uplink pilot block, i.e., it introduces no additional training overhead. The nonlinear least-squares estimator is more computationally demanding, but it uses all available summed phase-differences under the exact spherical-wave model and thereby offers the highest accuracy.

\section{System Model and Problem Formulation}

\subsection{System and Channel Model}
We consider a localization scenario with a single-antenna user and a UPA in a line-of-sight (LoS) environment. A Cartesian coordinate system, $Oxyz$, is centered at the UPA, which lies in the $Oxz$ plane. The UPA comprises $(2N+1) \times (2N+1)$ antenna elements with uniform inter-element spacing $d$ along both axes. The position vector of the antenna element at index $(n,m)$, where $n, m \in \{-N, \dots, N\}$, is $\boldsymbol{s}_{n,m} = [nd, 0, md]^T$.

The user is at an unknown position $\boldsymbol{r}$, described by spherical coordinates $(r, \theta, \phi)$, where $r = \|\boldsymbol{r}\|$ is the radial distance, $\theta$ is the azimuth angle, and $\phi$ is the zenith angle. In Cartesian coordinates, the user position is $\boldsymbol{r} = [r\sin\phi\cos\theta,\; r\sin\phi\sin\theta,\; r\cos\phi]^T$. Because the UPA lies in the $y=0$ plane, the measurements are symmetric with respect to reflection across this plane; thus only $|y|$ is identifiable from a single UPA. Throughout the paper we assume the user lies in the half-space $y\ge 0$ (equivalently, $\theta\in[0,\pi]$ for $\phi\in[0,\pi]$).

The exact distance between the user and the $(n,m)$-th antenna is given by \cite{liu2023}
\begin{align}\label{eq_diff_nm}
   & \|\boldsymbol{r} - \boldsymbol{s}_{n,m}\|\nonumber\\
    &= \sqrt{r^2 + d^2(n^2+m^2) - 2dnr\sin\phi\cos\theta - 2dmr\cos\phi}.
\end{align}
The channel coefficient $h_{n,m}$ between the user and the $(n,m)$-th antenna, based on the spherical-wave model, is
\begin{align}
    h_{n,m} = \frac{\sqrt{G_1(\boldsymbol{r},\boldsymbol{s}_{n,m}) G_2(\boldsymbol{r},\boldsymbol{s}_{n,m})}}{4 \pi \|\boldsymbol{r}-\boldsymbol{s}_{n,m}\|} e^{-j \frac{2\pi}{\lambda} \|\boldsymbol{r}-\boldsymbol{s}_{n,m}\|},
    \label{channel}
\end{align}
where $\lambda$ is the carrier wavelength, and $G_1(\cdot)$ and $G_2(\cdot)$ model the effective aperture and polarization losses, respectively.

\subsection{Problem Formulation}
During uplink training, the user transmits a sequence of $K$ known unit-modulus pilot symbols. After pilot removal, the effective signal received at the $(n,m)$-th antenna for the $k$-th pilot is
\begin{equation}
    y_{n,m,k} = \sqrt{P} h_{n,m} + w_{n,m,k},
\end{equation}
where $P$ is the average transmit power and $w_{n,m,k} \sim \mathcal{CN}(0, \sigma^2)$ is the Additive White Gaussian Noise (AWGN). We consider a single-user pilot interval. Hence, multi-user pilot reuse, pilot contamination, and user-separation issues are outside the scope of the present paper. We also assume coherent reception at the array with a common reference so that inter-element phase-differences are meaningful.

Our objective is to estimate the user's spherical coordinates $(r, \theta, \phi)$ from the received signals $\{y_{n,m,k}\}$. We aim to derive multiple closed-form expressions for these coordinates, assuming the gain functions $G_1(\cdot)$, $G_2(\cdot)$, and the noise variance $\sigma^2$ are unknown.

\section{Proposed Phase-Difference Localization Scheme}

\noindent\textbf{Main idea:} The proposed scheme operates in four steps. First, adjacent-antenna phase-differences are estimated along the two axes of the UPA. Second, these neighboring phase-differences are summed along rows and columns, which yields phase-differences between widely separated antennas without directly measuring potentially wrapped long-baseline phases. Third, three suitably selected summed relations are used to estimate $A=r^2$, $B=r\sin\phi\cos\theta$, and $C=r\cos\phi$, from which $r$, $\theta$, and $\phi$ follow directly. Finally, all available summed phase-differences are jointly exploited in a nonlinear least-squares refinement initialized by the closed-form estimate.

In this section, we present a method for deriving numerous closed-form expressions for the user coordinates $r$, $\theta$, and $\phi$. We also derive the number of candidate three-equation systems, which depends on the UPA size parameter $N$. We assume the UPA has access to $y_{n,m,k}$ for all $n,m \in \{-N,\ldots,N\}$ and $k \in \{1,\ldots,K\}$.

By averaging the received signal over $K$ pilot symbols, we obtain an estimate with reduced noise variance
\begin{equation} \label{ynm}
  y_{n,m} = \frac{1}{K}\sum_{k=1}^K y_{n,m,k} = \sqrt{P}h_{n,m} + w_{n,m},
\end{equation}
where the new noise term $w_{n,m}$ has zero mean and variance $\mathbb{E}(|w_{n,m}|^2) = \sigma^2/K$. The averaged received signal in \eqref{ynm} can be expressed equivalently as
\begin{align}
   y_{n,m} &= h_{n,m} \left(\sqrt{P} + \frac{w_{n,m}}{h_{n,m}} \right) \\
   &= |h_{n,m}| |z_{n,m}| \exp \left(
   j \left(
   -\frac{2\pi}{\lambda} \|\boldsymbol{r} - \boldsymbol{s}_{n,m}\| + \xi_{n,m}
   \right)
   \right),
\end{align}
where $z_{n,m} = \sqrt{P} + w_{n,m}/h_{n,m}$, and $\xi_{n,m} = \angle z_{n,m}$ denotes the phase-noise term.

Our method begins by measuring the phase-difference between signals at adjacent antennas. We define the measured phase-differences along the $x$ and $z$ axes as
\begin{align}
    \hat{\Phi}_{(n-1,m)}^{(n,m)} &= \angle\left(\frac{y_{n,m}}{y_{n-1,m}}\right) = \Phi_{(n-1,m)}^{(n,m)} + \xi_{n,m} - \xi_{n-1,m}, \\
    \hat{\Phi}_{(n,m-1)}^{(n,m)} &= \angle\left(\frac{y_{n,m}}{y_{n,m-1}}\right) = \Phi_{(n,m-1)}^{(n,m)} + \xi_{n,m} - \xi_{n,m-1},
\end{align}
where $\Phi_{(n-1,m)}^{(n,m)}$ and $\Phi_{(n,m-1)}^{(n,m)}$ are the corresponding noiseless phase-differences among neighboring antennas along the $x$ and $z$ axes, i.e.,
\begin{align}
   \Phi_{(n-1,m)}^{(n,m)}&= -\frac{2\pi}{\lambda} \left( \|\boldsymbol{r} - \boldsymbol{s}_{n,m}\| - \|\boldsymbol{r} - \boldsymbol{s}_{n-1,m}\| \right ),\nonumber \\
 \Phi_{(n,m-1)}^{(n,m)}&= -\frac{2\pi}{\lambda} \left( \|\boldsymbol{r} - \boldsymbol{s}_{n,m}\| - \|\boldsymbol{r} - \boldsymbol{s}_{n,m-1}\| \right).
\end{align}
From the triangle inequality, it follows that $\|\boldsymbol{r} - \boldsymbol{s}_{n,m}\| - \|\boldsymbol{r} - \boldsymbol{s}_{n-1,m}\| \in [-d, d]$ and $\|\boldsymbol{r} - \boldsymbol{s}_{n,m}\| - \|\boldsymbol{r} - \boldsymbol{s}_{n,m-1}\| \in [-d, d]$. Thus for $d \leq \lambda/2$, we ensure that $\Phi_{(n-1,m)}^{(n,m)} \in [-\pi, \pi]$ and $\Phi_{(n,m-1)}^{(n,m)} \in [-\pi, \pi]$. This condition prevents phase ambiguity in the noiseless adjacent-antenna phase-differences and allows us to avoid modulo-$2\pi$ operations, provided that the phase-noise is not excessively large.

Even when $d \le \lambda/2$, the measured adjacent phase-differences may still wrap under very low effective SNR because of the noise terms $\xi_{n,m}-\xi_{n-1,m}$ and $\xi_{n,m}-\xi_{n,m-1}$. Averaging over $K$ pilots reduces this risk by reducing the phase-noise variance.

Following \cite{n.zlat}, we sum the adjacent phase-differences along an axis to obtain the phase-difference between non-adjacent antennas while avoiding long-baseline phase wrapping on that axis. Specifically, for any integers $L < M$, we define the summed phase-differences along the $x$-axis as
\begin{align}
    \hat{S}_{(L,m)}^{(M,m)} &= \sum_{n=L+1}^M \hat{\Phi}_{(n-1,m)}^{(n,m)} =    S_{(L,m)}^{(M,m)} + \xi_{M,m} - \xi_{L,m}, \label{SLMm}
\end{align}
and along the $z$-axis as
\begin{align} \label{SLMn}
    \hat{S}_{(n,L)}^{(n,M)} &= \sum_{m=L+1}^M \hat{\Phi}_{(n,m-1)}^{(n,m)}  = S_{(n,L)}^{(n,M)}+ \xi_{n,M} - \xi_{n,L},
\end{align}
where $S_{(L,m)}^{(M,m)}$ and $S_{(n,L)}^{(n,M)} $ are the noiseless counterparts, given by
\begin{align}
    S_{(L,m)}^{(M,m)} &= -\frac{2\pi}{\lambda } \left( \|\boldsymbol{r} - \boldsymbol{s}_{M,m}\| - \|\boldsymbol{r} - \boldsymbol{s}_{L,m}\| \right),\label{SLMn_noiseless_2}\\
    S_{(n,L)}^{(n,M)}&= -\frac{2\pi}{\lambda } \left( \|\boldsymbol{r} - \boldsymbol{s}_{n,M}\| - \|\boldsymbol{r} - \boldsymbol{s}_{n,L}\| \right). \label{SLMn_noiseless_1}
\end{align}
The identities in \eqref{SLMn_noiseless_2}--\eqref{SLMn_noiseless_1} follow directly from telescoping the summed neighboring distance-differences. Crucially, this summation yields the phase-difference between widely separated antennas, a quantity that cannot in general be measured directly via $\angle(y_{n,M}/y_{n,L})$ because of phase wrapping.

Inserting \eqref{eq_diff_nm} into \eqref{SLMn_noiseless_2} for $(L,M)=(H_L,H_M)$, then inserting \eqref{eq_diff_nm} into \eqref{SLMn_noiseless_1} for $(L,M)=(V_L,V_M)$, and making the change of variables $A=r^2$, $B=r\sin\phi\cos\theta$ and $C=r\cos\phi$, we obtain
\begin{align}
     S_{(H_L,m)}^{(H_M,m)} &(A,B,C) = \nonumber\\
    &-\frac{2\pi}{\lambda } \Big( \sqrt{A + d^2(H_M^2+m^2) - 2BH_M d - 2Cmd} \nonumber \\
& - \sqrt{A + d^2(H_L^2+m^2) - 2B H_L d - 2Cmd} \Big), \label{SLMm_noiseless_yu}
\end{align}
\begin{align}
     S_{(n,V_L)}^{(n,V_M)}& (A,B,C) \nonumber\\
   &= -\frac{2\pi}{\lambda } \Big( \sqrt{A + d^2(n^2+V_M^2) - 2Bdn - 2C V_M d} \nonumber \\
& - \sqrt{A + d^2(n^2+V_L^2) - 2Bdn - 2C V_L d} \Big), \label{SLMn_noiseless_yu}
\end{align}
where $S_{(L,m)}^{(M,m)}$ and $S_{(n,L)}^{(n,M)}$ are viewed explicitly as functions of $A$, $B$, and $C$.

Equations \eqref{SLMm_noiseless_yu} and \eqref{SLMn_noiseless_yu} allow us to choose arbitrary triplets of $(H_L,H_M,m)$ and $(V_L,V_M,n)$, subject to $-N\leq H_L<H_M\leq N$, $-N\leq V_L<V_M\leq N$, and $-N\leq n,m\leq N$, and thereby form three nonlinear equations in $(A,B,C)$. If the corresponding noiseless sums were known, these equations would yield $A$, $B$, and $C$, and hence $r$, $\theta$, and $\phi$. In practice, however, only the noisy sums $\hat S_{(H_L,m)}^{(H_M,m)}$ and $\hat S_{(n,V_L)}^{(n,V_M)}$ in \eqref{SLMm} and \eqref{SLMn} are available. Replacing the noiseless sums in \eqref{SLMm_noiseless_yu} and \eqref{SLMn_noiseless_yu} with their measured counterparts yields the following approximate equations for estimating $A$, $B$, and $C$:
\begin{align}
    \hat S_{(H_L,m)}^{(H_M,m)} &\approx -\frac{2\pi}{\lambda } \Big( \sqrt{A + d^2(H_M^2+m^2) - 2BH_M d - 2Cmd} \nonumber \\
& - \sqrt{A + d^2(H_L^2+m^2) - 2B H_L d - 2Cmd} \Big), \label{SLMm_noiseless_yu-1}\\
    \hat S_{(n,V_L)}^{(n,V_M)}&\approx  -\frac{2\pi}{\lambda } \Big( \sqrt{A + d^2(n^2+V_M^2) - 2Bdn - 2C V_M d} \nonumber \\
& - \sqrt{A + d^2(n^2+V_L^2) - 2Bdn - 2C V_L d} \Big). \label{SLMn_noiseless_yu-1}
\end{align}

Any three summed phase-difference relations (horizontal/vertical) yield a three-equation nonlinear system in $(A,B,C)$, producing a candidate estimate. Let
\begin{equation*}
    Q = (2N+1)\binom{2N+1}{2}
\end{equation*}
denote the number of available summed phase-differences along one axis. Then the total number of three-equation systems that can be formed is
\begin{equation*}
    N_{\mathrm{sys}} = 2\binom{Q}{3} + 2\binom{Q}{2}Q.
\end{equation*}
This count is mainly of structural interest; In practice, we use either one carefully selected triplet of summed phase-difference relations or the all-data least-squares estimator below.

In general, solutions can be derived for $A$, $B$, and $C$ for almost all of the $N_{\mathrm{sys}}$ systems of three equations, except for the ones for which the choices of $(H_L,H_M, m)$ and/or $(V_L,V_M, n)$ lead to an ill-posed system of equations. In this paper, we are interested in finding the choices of $(H_L,H_M, m)$ and $(V_L,V_M, n)$ that lead to very simple expressions for $A$, $B$, and $C$, and thereby very simple expressions for $r$, $\phi$, and $\theta$. One such choice is $(H_L,H_M, m)=(-N,N,0)$, $(V_L,V_M, n)=(-N,0,0)$ and $(V_L,V_M, n)=(0,N,0)$, which leads to
\begin{align}
 A &=  \left(\frac{\Delta_{V_1}^{\,2} + \Delta_{V_2}^{\,2} - 2 d^2 N^2}{\,2(\Delta_{V_1} - \Delta_{V_2})\,}\right)^{\!2}, \label{eq_A}\\
 C&=  \frac{2\Delta_{V_1} \sqrt{A} + d^2 N^2 - \Delta_{V_1}^{\,2}}{\,2 d N\,}, \label{eq_C} \\
 B &=   \frac{ \Delta_H \sqrt{4A+4 d^2 N^2-\Delta_H^2 }}{\,4 d N\,}.  \label{eq_B}
\end{align}
where
\begin{align}\label{eq_measurem}
 \Delta_{V_1} := \frac{\lambda \hat S_{(0,\,0)}^{(0,\, N)}}{2 \pi},\; 
 \Delta_{V_2} := \frac{\lambda \hat S_{(0,\,-N)}^{(0,\, 0)}}{2 \pi}, \nonumber\\
 \text{and } 
 \Delta_H := \frac{\lambda \hat S_{(-N,\,0)}^{(N,\, 0)}}{2 \pi}.
\end{align}
The corresponding algebraic elimination is omitted here for brevity.

From $A$, $B$, and $C$ in \eqref{eq_A}--\eqref{eq_B}, we can obtain estimates for $r$, $\theta$, and $\phi$ as
\begin{align}
   r &= \sqrt{A}, \label{eq_r}\\
    \theta &=  \arccos\!\Big(\frac{B}{\sqrt{A-C^2}}\Big)\in[0,\pi] , \label{eq_theta}\\
    \phi &= \arccos\!\Big(\frac{C}{\sqrt{A}}\Big)\in[0,\pi]. \label{eq_phi}
\end{align}

Note that $(A,B,C)$ correspond to $(r^2,x,z)$ in Cartesian coordinates, i.e., $x=B$ and $z=C$. Under the half-space assumption $y\ge 0$, we can recover the full 3D position estimate as
\begin{equation}
    \hat{\boldsymbol{r}} = \big[\hat B,\; \sqrt{\hat A-\hat B^2-\hat C^2},\; \hat C \big]^T,
\end{equation}
whenever $\hat A-\hat B^2-\hat C^2\ge 0$.

The above closed-form estimator uses only one triplet of phase-difference sums. To exploit all available phase-difference sums, we also consider the following nonlinear least-squares problem:
\begin{align}\label{eq_OP}
&(A,B,C) =  \underset{\substack{A,B,C\\ A>0,\,A\geq B^2+C^2}}{\arg\min} \Bigg[   \sum_{m=-N}^N \sum_{H_L=-N}^N \sum_{H_M=H_L+1}^N \nonumber\\
 & \qquad\qquad\qquad\qquad\qquad
   \Big(   \hat S_{(H_L,m)}^{(H_M,m)}   - S_{(H_L,m)}^{(H_M,m)}(A,B,C) \Big)^2
  \nonumber\\
  &+
  \sum_{n=-N}^N \sum_{V_L=-N}^N \sum_{V_M=V_L+1}^N  \left(  \hat S_{(n,V_L)}^{(n,V_M)} - S_{(n,V_L)}^{(n,V_M)}(A,B,C) \right)^2
  \Bigg],
\end{align}
where $\hat S_{(H_L,m)}^{(H_M,m)}$ and $\hat S_{(n,V_L)}^{(n,V_M)}$ are obtained from \eqref{SLMm} and \eqref{SLMn}, while $S_{(H_L,m)}^{(H_M,m)}(A,B,C)$ and $S_{(n,V_L)}^{(n,V_M)}(A,B,C)$ are given by \eqref{SLMm_noiseless_yu} and \eqref{SLMn_noiseless_yu}, respectively. The constraints $A>0$ and $A\geq B^2+C^2$ ensure a valid 3D geometry. The problem in \eqref{eq_OP} is a standard nonlinear least-squares problem and can be solved numerically, for example via the Levenberg--Marquardt algorithm. A convenient initialization is provided by the closed-form estimate in \eqref{eq_A}--\eqref{eq_B}.

Computationally, averaging the pilots and forming all adjacent phase-differences over the $(2N+1)^2$ antennas require $O(K(2N+1)^2)$ and $O((2N+1)^2)$ operations, respectively. All row/column summed phase-differences can then be generated using prefix sums in $O(N^3)$. Thus, the representative closed-form estimator is single-shot after preprocessing, whereas the nonlinear least-squares estimator has a per-iteration cost of $O(N^3)$ because it uses all available summed phase-differences. The tradeoff is therefore clear: the closed-form estimator offers minimal computational load, while the least-squares estimator offers the best accuracy at the cost of numerical iterations.

\section{Numerical Results}
We evaluate the proposed localization scheme for $\lambda=0.01$~m, $d=\lambda/2$, transmit power $P=23$~dBm, noise power $\sigma^2=-114$~dBm, and isotropic antennas, i.e., $G_1(\boldsymbol{r},\boldsymbol{s}_{n,m})=1$ and $G_2(\boldsymbol{r},\boldsymbol{s}_{n,m})=\lambda^2/(4\pi)$. Each point is averaged over $500$ Monte Carlo realizations, with $\theta=\pi/6$ and $\phi=\pi/4$. We compare the proposed closed-form estimator, the proposed all-sum least-squares (LS) estimator initialized by the closed-form estimate, APLE, E-APLE \cite{baseline}, and the 3D position CRB. The APLE/E-APLE hyperparameters are fixed over all sweeps.

The reported metric is the absolute 3D position RMSE in meters, computed in Cartesian coordinates before averaging:
\begin{align}
\mathrm{RMSE}_{\mathrm{3D}}
&=\sqrt{\frac{1}{M}\sum_{i=1}^{M}
\|\hat{\boldsymbol p}^{(i)}-\boldsymbol p\|^2}, \label{eq:rmse_3d}\\
\boldsymbol p
&=[r\sin\phi\cos\theta,\; r\sin\phi\sin\theta,\; r\cos\phi]^T. \nonumber
\end{align}
For the proposed estimators, $\hat{\boldsymbol p}$ is reconstructed from $(\hat r,\hat\theta,\hat\phi)$, or equivalently from $(\hat A,\hat B,\hat C)$ as $\hat{\boldsymbol p}=[\hat B,\sqrt{\hat A-\hat B^2-\hat C^2},\hat C]^T$ under the assumed $y\ge 0$ half-space. The APLE/E-APLE outputs are mapped to the same coordinate convention before computing \eqref{eq:rmse_3d}.

To explicitly include a classical near-field case, we use the standard Fraunhofer distance $R_F=2D^2/\lambda$. For the considered $(2N+1)\times(2N+1)$ UPA, the side and diagonal apertures are $D_{\mathrm{s}}=2Nd$ and $D_{\mathrm{d}}=2\sqrt{2}Nd$, respectively. Since $d=\lambda/2$,
\begin{equation}
R_{F,\mathrm{s}}=2N^2\lambda,
\qquad
R_{F,\mathrm{d}}=4N^2\lambda.
\label{eq:fraunhofer}
\end{equation}
For $N=20$, \eqref{eq:fraunhofer} gives $R_{F,\mathrm{s}}=8$~m and $R_{F,\mathrm{d}}=16$~m; hence the $r=5$~m user is in the classical near-field region under either aperture convention, whereas the $r=50$~m user is in the far-field region.

\begin{figure}[!t]
    \centering
    \includegraphics[width=0.95\columnwidth]{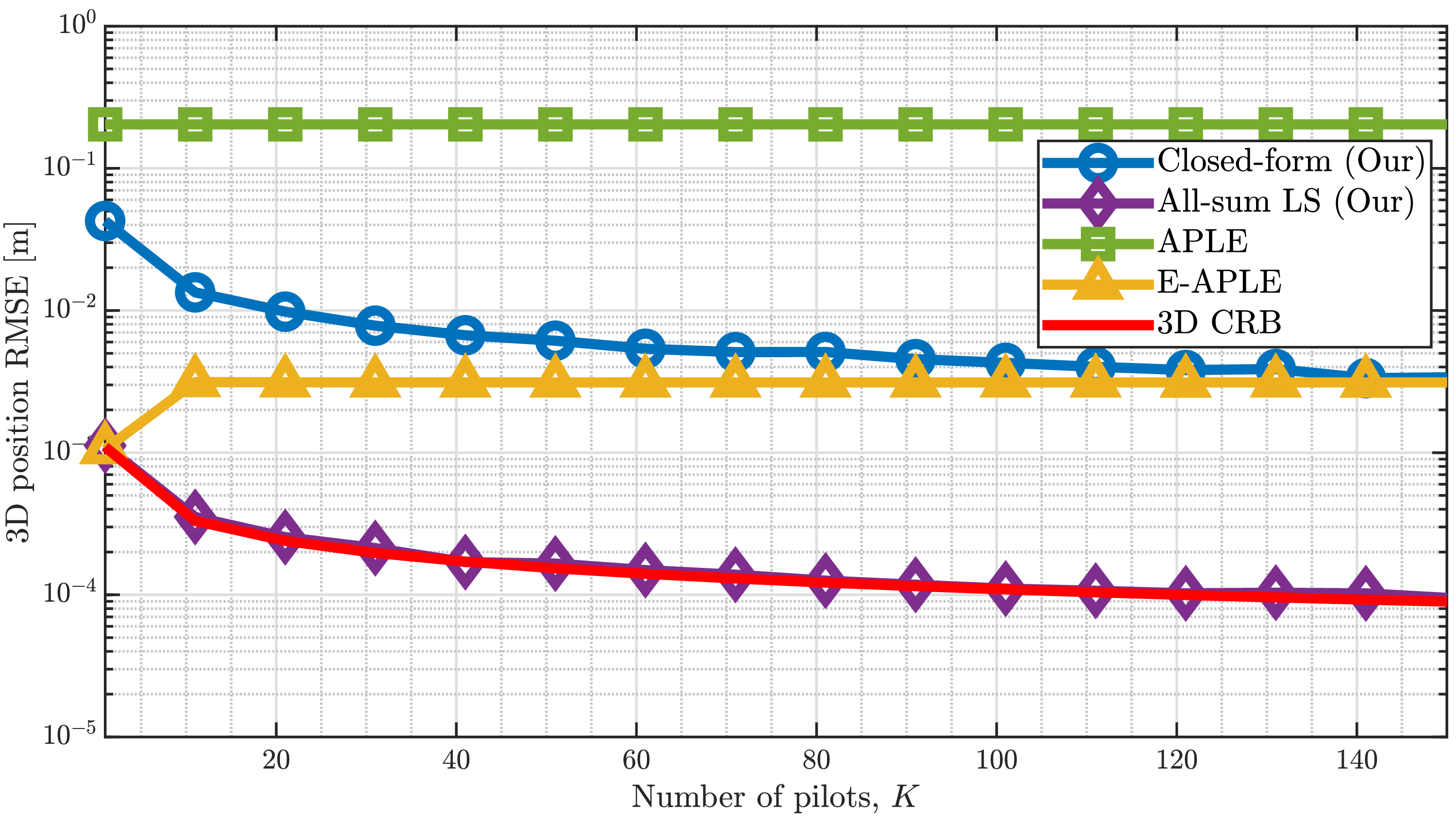}
    \caption{Absolute 3D position RMSE versus the number of uplink pilots $K$ for the classical near-field user at $r=5$~m and $N=20$.  The angles are $\theta=\pi/6$ and $\phi=\pi/4$.}
    \label{fig:posK_nf}
\end{figure}

\begin{figure}[!t]
    \centering
    \includegraphics[width=0.95\columnwidth]{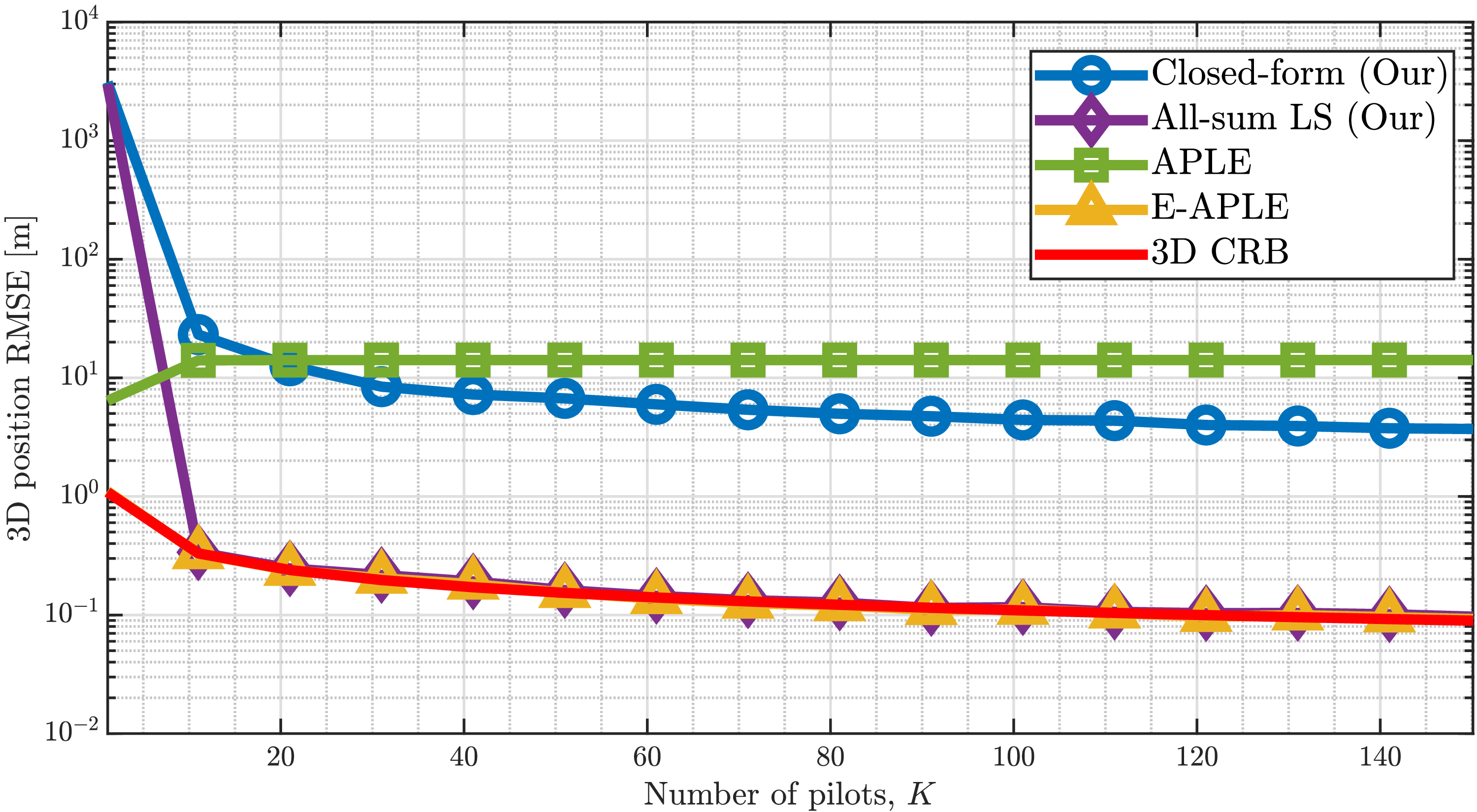}
    \caption{Absolute 3D position RMSE versus the number of uplink pilots $K$ for the far-field comparison user at $r=50$~m and $N=20$.   The angles are $\theta=\pi/6$ and $\phi=\pi/4$.}
    \label{fig:posK_ff}
\end{figure}

\Cref{fig:posK_nf,fig:posK_ff} show the absolute 3D position RMSE versus the number of pilots $K$ for $r=5$~m and $r=50$~m, respectively. In the classical near-field case, i.e., for $r=5$~m, \Cref{fig:posK_nf} shows that the proposed all-sum LS estimator closely follows the 3D position CRB over the plotted range, while the closed-form estimator has a larger but decreasing error because it uses only three selected phase-sum relations. The APLE baseline exhibits an error floor, and E-APLE remains above the proposed all-sum LS estimator for moderate and large $K$. In the far-field comparison, i.e., for $r=50$~m, \Cref{fig:posK_ff} shows that the weaker spherical curvature makes reliable initialization of the proposed all-sum LS estimator more difficult at very small $K$; however, once $K$ is moderately large, both the proposed all-sum LS estimator and E-APLE approach the 3D position CRB. These figures highlight a practical drawback of the APLE/E-APLE pipeline: its final accuracy can be sensitive to initialization quality and hyperparameter choices, whereas the proposed closed-form initialization is deterministic and requires no algorithmic fine-tuning.

\begin{figure}[!t]
    \centering
    \includegraphics[width=0.95\columnwidth]{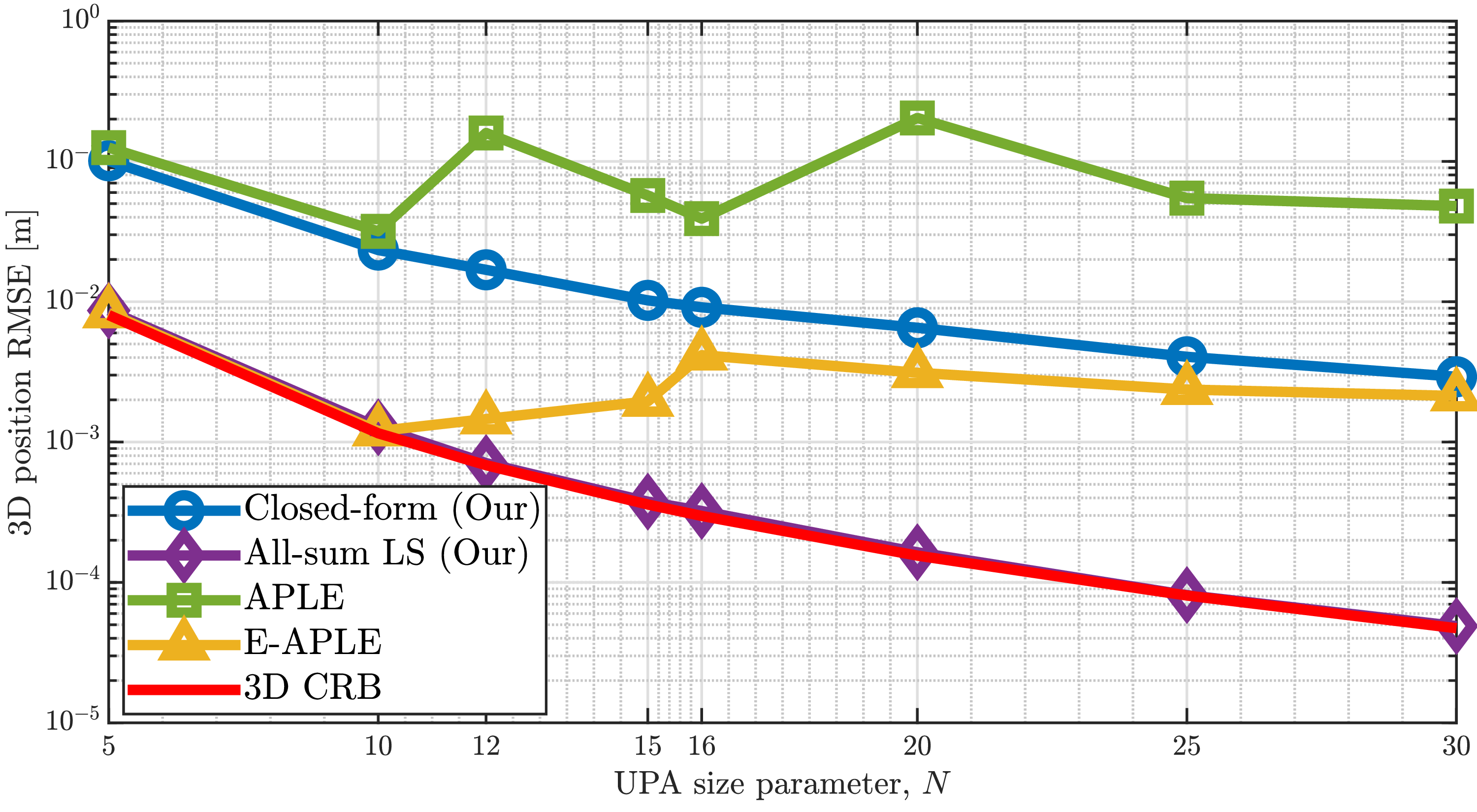}
    \caption{Absolute 3D position RMSE versus the UPA size parameter $N$ for the $r=5$~m user and $K=50$ pilots.   The angles are $\theta=\pi/6$ and $\phi=\pi/4$.}
    \label{fig:posN_nf}
\end{figure}

\begin{figure}[!t]
    \centering
    \includegraphics[width=0.95\columnwidth]{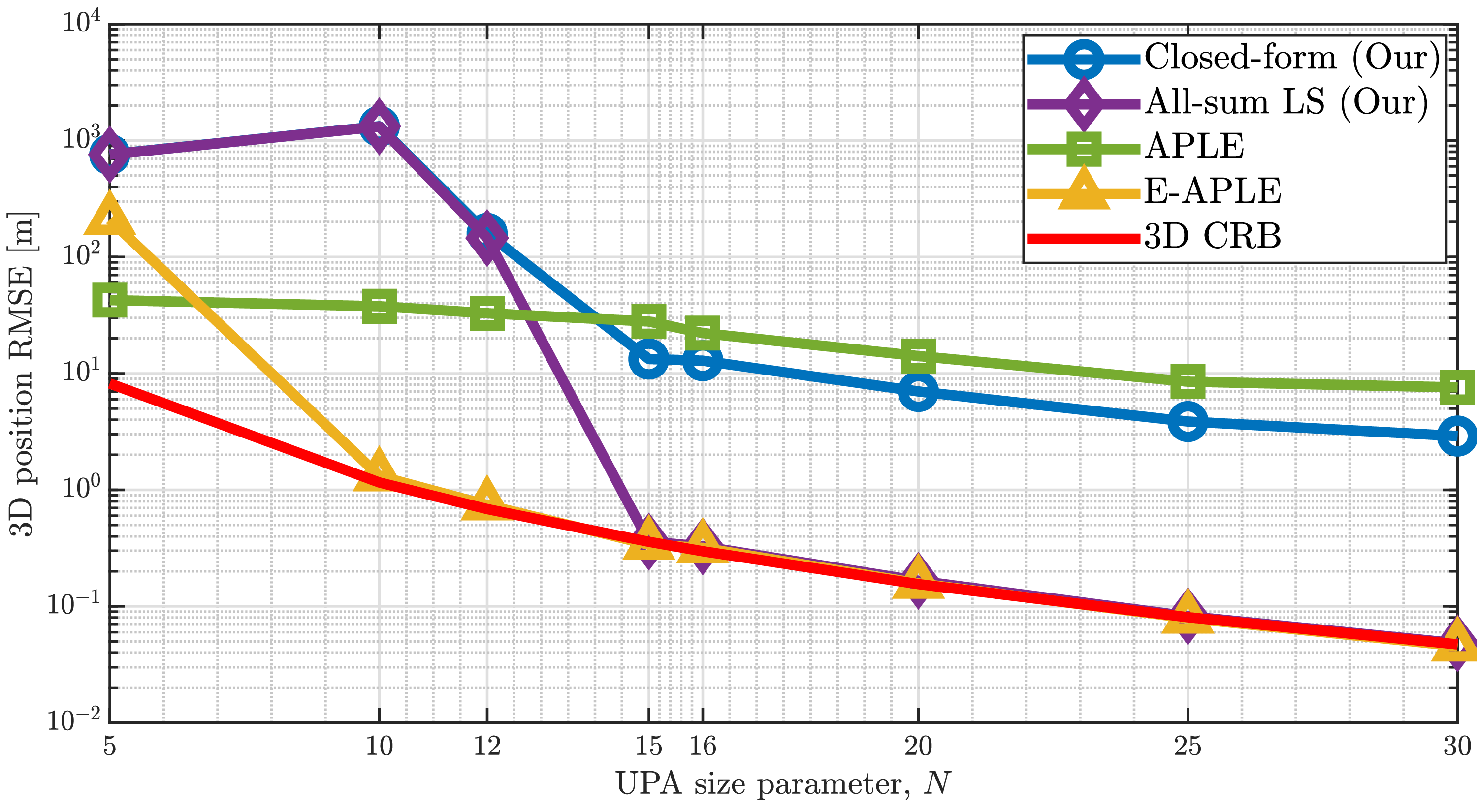}
    \caption{Absolute 3D position RMSE versus the UPA size parameter $N$ for the far-field comparison user at $r=50$~m and $K=50$ pilots.   The angles are $\theta=\pi/6$ and $\phi=\pi/4$.}
    \label{fig:posN_ff}
\end{figure}

\Cref{fig:posN_nf,fig:posN_ff} show the absolute 3D position RMSE versus $N$ for $K=50$ pilots and $r=5$~m and $r=50$~m, respectively. For $r=5$~m, the user is in the classical near-field region for $N\ge 12$ under the diagonal-aperture convention and for $N\ge 16$ under the side-aperture convention. The all-sum LS estimator tracks the 3D position CRB over the plotted range of $N$. For $r=50$~m, the user remains outside the classical near-field region over the plotted range. At small apertures, the weak range information can lead to poor LS initialization, whereas sufficiently large apertures allow the proposed all-sum LS estimator to approach the 3D position CRB. As expected, the RMSE of the proposed methods decreases as $N$ increases. This is not always the case for the baseline schemes, whose hyperparameters may need to be adjusted across different values of $N$ and different user distances to achieve their best performance.

\section{Conclusion}
We proposed a 3D UPA localization method under the exact spherical-wave model based on summed horizontal and vertical phase-differences. The method yields a representative closed-form estimator and an all-sum  LS estimator. The numerical results show that the proposed all-sum LS estimator closely follows the 3D position CRB in this near-field case and approaches the CRB in the far-field comparison when the number of pilots or the aperture size is sufficiently large.

\bibliographystyle{IEEEtran}
\bibliography{biblio}

\end{document}